\documentclass[pra,twocolumn,showpacs,preprintnumbers,
amsmath,amssymb,tightenlines,epsfig,floatfix]{revtex4}

\usepackage{bm}
\usepackage{graphicx}
\usepackage{color}

\newcommand{\beq}{\begin{equation}}
\newcommand{\eeq}{\end{equation}}

\newcommand{\be}{\begin{equation}}
\newcommand{\ee}{\end{equation}}
\newcommand{\bi}{\begin{itemize}}
\newcommand{\ei}{\end{itemize}}
\newcommand{\ben}{\begin{enumerate}}
\newcommand{\een}{\end{enumerate}}

\newcommand{\pol}{\hat{\bf e}}
\newcommand{\rv}{{\bf r}}

\newcommand{\kv}{{\bf k}}

\newcommand{\bea}{\begin{eqnarray}}
\newcommand{\eea}{\end{eqnarray}}

\renewcommand{\>}{\rangle}

\newcommand{\commentout}[1]{{}}
\newcommand{\half}{\hbox{$1\over2$}}
\newcommand{\eq}[1]{Eq.~\eqref{#1}}

\definecolor{black}{rgb}{0,0,0}

\definecolor{blue}{rgb}{0,0,1}

\definecolor{green}{rgb}{0,1,0}

\definecolor{red}{rgb}{1,0,0}

\begin{document}
\title{Optical Kagom\'e lattice for ultra-cold atoms with nearest neighbor interactions}
\author{J. Ruostekoski}
\affiliation{School of Mathematics, University of Southampton,
Southampton, SO17 1BJ, UK}

\begin{abstract}
We propose a scheme to implement an optical Kagom\'e lattice for ultra-cold atoms 
with controllable $s$-wave interactions between nearest neighbor sites and a gauge potential. 
The atoms occupy three 
different internal atomic levels with electromagnetically-induced coupling between 
the levels. We show that by appropriately shifting the triangular lattice potentials,
experienced by atoms in different levels, the Kagom\'e lattice can be realized using
only two standing waves, generating a highly frustrated quantum system for the atoms.  
\end{abstract}
\date{\today}
\maketitle

Ultra-cold atomic gases in periodic optical lattice potentials can form a highly-controllable
quantum many-particle system that has demonstrated interesting analogies to crystal lattice
systems of strongly interacting electrons. Among such experimental developments are the Mott 
insulator states of both bosonic \cite{GRE02a} and fermionic atoms \cite{JOR08,SCH08}, 
fermionic superfluidity \cite{CHI06}, atom transport \cite{FER04,KIN06,MUN07}, and super-exchange 
correlations in a collection of double wells \cite{superexc}. The rapid progress has 
generated interest in engineering ultra-cold atomic lattice systems
that could provide clean realizations of Hubbard models of strongly-correlated crystal
lattice systems with the potential use of atoms as quantum simulators 
of some unresolved models, e.g., in high-$T_c$ superconductivity \cite{hightc}.  

One of the challenging problems in the theory of strongly correlated systems has been to 
characterize the phase diagrams of two-dimensional (2D) and 3D Hubbard models of geometrically
frustrated lattices, with competing interactions resulting in highly degenerate
ground states. Of particular interest are corner-sharing networks of complete graphs, such
as Kagom\'e and pyrochlore/checkerboard lattices.  It was recently proposed that
frustration in pyrochlore and diamond lattices could generate fractional charges
in the presence of nearest-neighbor (nn) repulsion \cite{FULDE}, with similar excitations
potentially existing in a Kagom\'e lattice. Moreover, spin-1/2 nn Heisenberg antiferromagnet
on a Kagom\'e lattice provides arguably the most promising candidate for a quantum 
spin liquid, with the Kagom\'e lattice exhibiting a higher degree of frustration than, e.g.,
a triangular lattice \cite{kagomereview}, and it has recently attracted considerable interest 
in {\it Volborthite} and {\it Herbertsmithite} combounds. Despite extensive theoretical
effort the true nature of the ground state of the Kagom\'e system has been evasive \cite{spinkagome}.
In addition, Kagom\'e systems can exhibit, e.g., kinetic ferromagnetism \cite{POL08} and 
production of trimerized and ideal Kagom\'e lattices for ultra-cold atoms has also started
attracting theoretical interest \cite{SAN04}.

In this paper we propose a constructed 2D optical Kagom\'e lattice system of neutral atoms
with collisional interactions between nn
sites. The hopping between the nn sites is induced by electromagnetic (em)
transitions between three internal atomic states. Due to the spin-dependent
lattice system, the Kagom\'e lattice can be prepared with only two optical standing 
waves (SWs), as compared to six SWs in previous proposals \cite{SAN04}, with
the additional advantage of controllable non-local two-body interactions between different sites
and the possibility for the creation of Abelian and non-Abelian guage potentials. 
The strength of the em-driven hopping can be tuned over a wide range of values with 
respect to tunneling between more distant sites and the nn collisions.  Spin-dependent 
lattices where the different atomic spin components were moved around independently was
experimentally created using $^{87}$Rb atoms \cite{MAN03}. Alkaline-earth-metal atoms and rare-earth 
metals with narrow optical resonances (e.g.\ Sr, Yb)
are particularly suitable for realizing spin-dependent optical lattices, 
because of slow loss rates due to spontaneous emission \cite{yb}.  
\begin{figure}
\includegraphics[width=0.48\columnwidth]{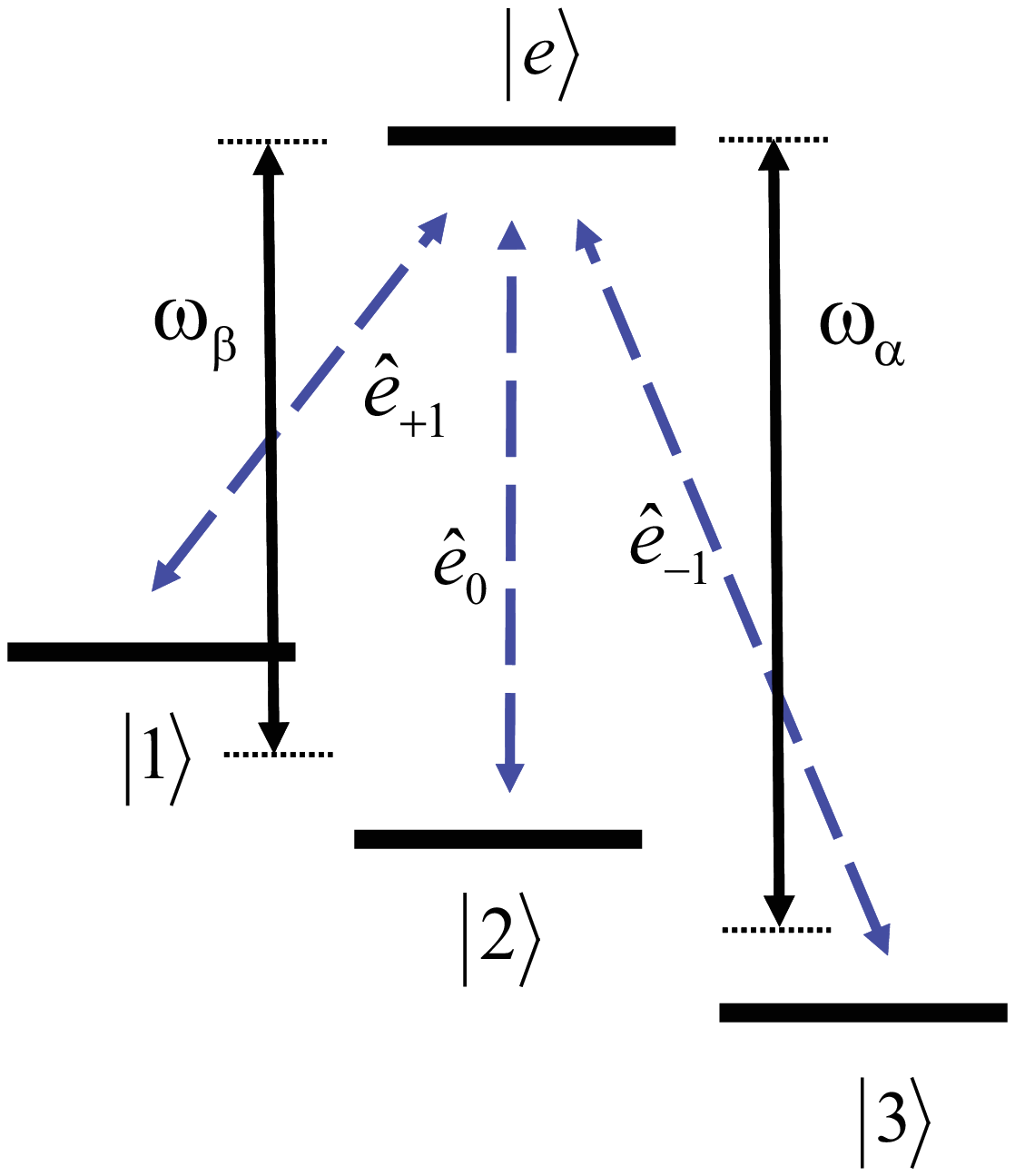}
\includegraphics[width=0.48\columnwidth]{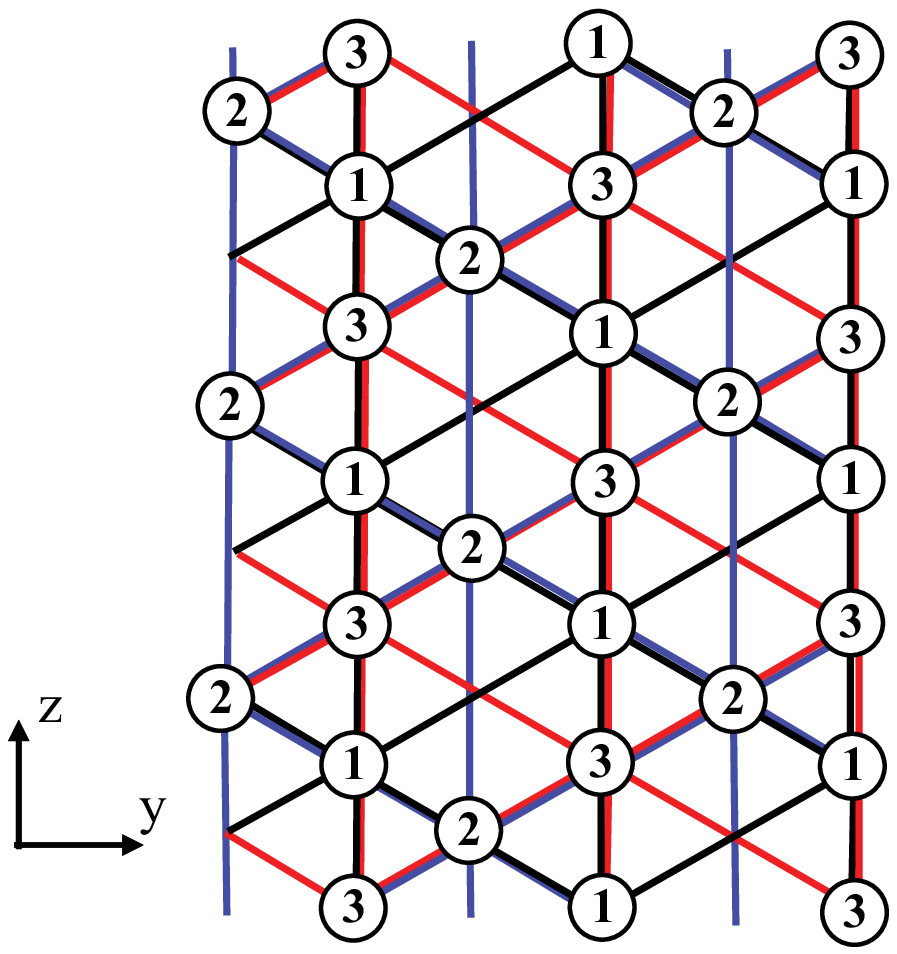}
\vspace{-3mm} \caption{Left: In the Kagom\'e lattice
the atoms occupy three electronic ground states with a common excited state.
The frequencies of the two lattice lasers are tuned between the atomic resonance frequencies
in such a way that $\omega_3>\omega_\alpha>\omega_{2}>\omega_\beta>\omega_1$. The polarization
components of the lattice lasers that couple to each individual transition are indicated.
Right: Atoms in each of the three internal levels $|j\>$ ($j=1,2,3$) experience an equilateral 
triangular lattice potential. These lattices are shifted with respect to each other, so that
the lattice site occupied by the atoms in $|2\>$ ($|3\>$) is at the midpoint of the triangle side of
$|1\>$ defined by the unit vector $\kv_\beta/k=(\pol_y - \sqrt{3}\pol_z)/2$ ($\pol_z$).
Each site has four nn sites,
so that, e.g., the site of $|1\>$ has two nn sites of $|2\>$ and two of
$|3\>$.} \label{fig1}
\end{figure}

We consider ultra-cold (bosonic or fermionic) atoms occupying three internal sublevels 
$|j\>$ ($j=1,2,3$) 
of the same atom that are coupled by em transitions. The atom dynamics is assumed
to be restricted to a 2D layer in the $yz$ plane due to a tightly confined 2D 
pancake-shaped trap. Such a confinement
may be experimentally created, e.g., using magnetic or optical fields \cite{KT,Smith}. 
Along the 2D trap each species
experiences a triangular optical lattice potential, generated by two SWs with wavevectors $\kv_{\alpha,\beta}=k(\pol_y \pm \sqrt{3}\pol_z)/2$. 
The atom-lattice model is based on an atomic (single-band) Hubbard Hamiltonian \cite{JAK98} where 
the atoms occupy the lowest mode of each lattice site. 
We show that by appropriately tuning the frequencies of the lattice lasers,
the lattice potentials of the three species can be shifted in a triangular
shape to form a Kagom\'e lattice pattern. Then the four nn sites are occupied
by the atoms in the other two sublevels and the hopping of the atoms between adjacent
sites, with amplitude $\kappa_{jk}$, only occurs as a result of driving by coherent em 
fields that change the internal atomic level. The hopping to the nearest 
sites occupied by the atoms in the same sublevel results from tunneling 
between the sites and, in sufficiently deep
lattices, it may be suppressed. Moreover, the specialty of 
the proposed scheme is that by adjusting the overlap 
of the nn lattice site wave functions (Wannier functions) we can prepare 
a lattice system with a non-negligible, controllable $s$-wave interaction between adjacent sites 
$U^{\rm nn}_{jk}$ ($j\neq k$), with the both
limits $U^{\rm nn}_{jk}\gg\kappa_{jk}$ and $U^{\rm nn}_{jk}\ll\kappa_{jk}$ achievable,
providing a frustrated quantum system with nn interactions. 

As a realization of a frustrated Kagom\'e
lattice with desired interactions, we consider a tripod four-level scheme; Fig.~\ref{fig1}. 
The atoms in the three electronic ground states are coupled to a common electronically excited
level $|e\>$ by different resonant transition frequencies $\omega_j$ ($\omega_3>\omega_2>\omega_1$).  
The lattice laser $\alpha$, with the frequency $\omega_\alpha$, is blue-detuned with respect to
two of the transitions $\delta_{\alpha,j}=\omega_j-\omega_\alpha<0$, for $j=1,2$, and red-detuned
with respect to one $\delta_{\alpha,3}>0$. Then the atoms in the states $|1\>$, $|2\>$ ($|3\>$) 
are attracted towards low-intensity (high-intensity) regions of $\alpha$.
The lattice laser $\beta$, with the frequency $\omega_\beta$, 
is blue-detuned from one of the transitions $\delta_{\beta,1}<0$ and red-detuned from the others.
For simplicity, we assume $|e\>=|e,m\>$, $|1\>=|1,m-1\>$, $|2\>=|2,m\>$,
and $|3\>=|3,m+1\>$, so that the transitions $|1\>\rightarrow |e\>$, $|3\>\rightarrow |e\>$, and
$|2\>\rightarrow |e\>$ have dipole matrix elements $\mathbf{d}_{e1} = \langle e|\mathbf{d}|1\rangle
=\mathfrak{D}\langle e|1,1 \rangle\hat{\mathbf{e}}_{+1}^{*}$, $\mathbf{d}_{e3} 
=\mathfrak{D}\langle e|-1,3\rangle\hat{\mathbf{e}}_{-1}^{*}$, and $\mathbf{d}_{e2} 
=\mathfrak{D}\langle e|0,2\rangle\hat{\mathbf{e}}_{0}^{*}$, coupling to the light with polarizations
$\sigma^+$, $\sigma^-$, and $\pol_0$, respectively. Here $\mathfrak{D}$ is the reduced dipole
matrix element, $\langle e|\sigma g\rangle $ are the Clebsch-Gordan
coefficients, and we use $\pol_{+1}=-(\pol_x+i\pol_y)/\sqrt{2}$, $\pol_{-1}=(\pol_x-i\pol_y)/\sqrt{2}$,
and $\pol_0=\pol_z$. 
We assume the polarizations of the two SWs to be orthogonal $\pol_\alpha\cdot\pol_\beta^*=0$, so
that the lattice potentials read $V_j=V_j^\alpha+V_j^\beta$ with 
\begin{align}
\label{potential}
V^\alpha_j= & s_{j\alpha} E_r\sin^2[k(y+
\sqrt{3}z)/2+\varphi_j],\nonumber\\
V_j^\beta= & s_{j\beta} E_r \sin^2[k(y-\sqrt{3}z)/2
+\eta_j]\,, 
\end{align}
$\varphi_1=\eta_1=\varphi_3=0$, $\varphi_2=\eta_2=\eta_3=\pi/2$. Here $s_{j\eta}\propto 
|(\mathbf{d}_{ej}\cdot \pol_\eta)^2/\delta_{\eta,j}|$ ($\eta=\alpha,\beta$) denotes
the lattice strength in the lattice photon recoil energy units $E_r=\hbar^2 k^2/2m$, depending on the light polarization, atomic sublevel, and the detuning. To produce three triangular lattices the lasers need to couple
simultaneously to all the three transitions, so that $s_{j\alpha}$ and $s_{j\beta}$ are non-vanishing for all $j$. This can be obtained by choosing the polarization vectors for the SWs so that $\pol_\eta\cdot\pol_{\pm1,0}^*\neq0$
together with $\pol_\eta\cdot\kv_\eta=0$ ($\eta=\alpha,\beta$) and $\pol_\alpha\cdot\pol_\beta^*=0$.

We will next demonstrate an example to show that such a solution can be found for the polarization
vectors and how selecting the lattice light polarization can also be used to
control the relative strengths of the three triangular lattices. This can be especially useful
if the absolute values of the detunings are very different. We consider the level scheme shown in Fig.~\ref{fig1} and, for simplicity, assume that $|\delta_{\alpha,3}|= |\delta_{\alpha,2}| =3 |\delta_{\alpha,1}|= 3 |\delta_{\beta,3}|= |\delta_{\beta,2}| = |\delta_{\beta,1}|$ 
(indicating that $\omega_1-\omega_2=\omega_2-\omega_3$ and that the lasers are detuned exactly at the midpoint
between the nearest transitions) and that the Clebsch-Gordan coefficients are equal. The polarization vectors (orthogonal to the corresponding wavevectors $\kv_j$) for the SWs are 
$\pol_\alpha = \mathfrak{a}_\alpha (-\sqrt{3}\pol_y+\pol_z)/2 +
\mathfrak{b}_\alpha \pol_x$ and $\pol_\beta = \mathfrak{a}_\beta (\sqrt{3}\pol_y+\pol_z)/2 +
\mathfrak{b}_\beta \pol_x$. Here we choose the complex coefficients $\mathfrak{a}_{\alpha,\beta}$ and
$\mathfrak{b}_{\alpha,\beta}$ in such a way that $\pol_\alpha\cdot\pol_\beta^*=0$ and $|\pol_{-1}^*\cdot \pol_\alpha|^2
= 3|\pol_{+1}^*\cdot \pol_\alpha|^2$ and $|\pol_{+1}^*\cdot \pol_\beta|^2
= 3|\pol_{-1}^*\cdot \pol_\beta|^2$. The last two conditions ensure that we have $s_{1\eta}=s_{3\eta}$, compensating for the different values of the detunings. A straightforward algebra yields solutions for
$\mathfrak{a}_{\alpha,\beta}$ and $\mathfrak{b}_{\alpha,\beta}$ with $s_{2\eta}=4s_{3\eta}/5$ ($\eta=\alpha,\beta$), resulting in only very small differences between the lattice strengths. 
We also find that the two SWs in \eq{potential} have
equal amplitudes for each level, i.e., $s_{j\alpha}=s_{j\beta}$ for all $j$. 

For each species the lattice potential $V_j$, in \eq{potential} is an equilateral triangle 
with the side of length $2d$, where $d=\pi/\sqrt{3}k$ denotes the
nn separation; Fig.~\ref{fig1}. The minima of the potential are at the corners of the triangles which 
for $|1\>$ are at $(y,z)=
[(n+m)\pi/k, (n-m)d]$, for $|2\>$ at $[(n+m+1/2)\pi/k, (n-m-1/2)d]$,
and for $|3\>$ at $[(n+m+1)\pi/k, (n-m)d]$, with $n$, $m$ integers. The triangular lattices 
of $|2\>$ and $|3\>$ are
shifted to coincide with the side midpoints of the triangular lattice of $|1\>$. 
The combined system of interlaced triangular lattices forms a Kagom\'e lattice, so that each site has four
nn sites which are occupied by the two other species (two of each). 

In order to estimate the relative strengths of the different terms in the lattice Hamiltonian
we evaluate the corresponding integral representations. The direct tunneling amplitude,
where atoms remain in the same hyperfine level during the hopping process, reads
$J^{\bf p}_b\simeq-\int dy\,dz\,(-{\hbar^2\over 2m}
\phi_{bn}^*\nabla_\parallel^2 \phi_{bn'}+\phi_{bn}^* V_b
\phi_{bn'})>0$,
where $\nabla_\parallel^2=\partial_y^2+\partial_z^2$ and $V_b$ is given by \eq{potential}.
The Wannier functions for atoms in $|b\>$ at site $n$ are $\phi_{bn}(y,z)$ and may be 
approximated by the ground state
harmonic oscillator wave function with the trap frequencies  $\omega_y\simeq \sqrt{2s} E_r/\hbar$ and $\omega_z\simeq \sqrt{6s} E_r/\hbar$ which are obtained by expanding the optical
potential at the lattice site minimum \cite{JAK98}. The site $n'$ here refers to the 
nearest site to $n$ occupied by the
same atomic hyperfine level along the direction of ${\bf p}$, where ${\bf p}$ takes the values of
the three triangle sides: $\kv_{\alpha,\beta}, \hat{\bf z}$. Here $J^{\kv_\alpha}_b= J^{\kv_\beta}_b$ by
symmetry. For $J^{\kv_\alpha}_b$ and $J^{\hat{\bf z}}_b$ we have $\phi_{b,n'}(y,z) = \phi_{b,n}(y+\sqrt{3}d,z+d)$  and $\phi_{b,n'}(y,z) = \phi_{b,n}(y,z+2d)$, respectively. 
Due to the anisotropy of the individual lattice site wavefunctions, the 
hopping amplitudes along the $z$ direction differ slightly from those along the direction 
of the two SWs.  Although a more rigorous calculation of the hopping amplitudes would involve a full
band-structure calculation, here it is sufficient to provide order-of-magnitude analytic estimates using
the Gaussian approximations to $\phi_{bn}$. 

The em field changes the internal level of the atom.
Because the lattice site minima of the different 
sublevels are shifted with respect to each other, the atoms
simultaneously also undergo spatial hopping along the lattice. 
The hopping amplitude reads
\begin{equation}
\kappa_{bc}=\int d^3 r\, \phi^*_{bn} \hbar {\cal R}_{bc}\phi_{cn'}\,, \label{kappa}
\end{equation}
where ${\cal R}_{bc}$ denotes the effective Rabi frequency for
the transition between the levels $|b\>$ and $|c\>$ and $n'$ refers
to the nn site of species $|c\>$ to the site $n$ occupied by the species $|b\>$,
with $\kappa_{bc}$ proportional 
to the spatial overlap between the atoms in the nn sites.
As shown in Fig.~\ref{fig1}, $\kappa_{12}$, $\kappa_{23}$, and $\kappa_{13}$ represent 
hopping along the directions of $\kv_\beta$, $\kv_\alpha$, and $\hat{\bf z}$, respectively. 
As for the direct tunneling
we then have $\kappa_{12}= \kappa_{23}$, but $\kappa_{13}$ is not exactly
equal. Note that $\kappa_{bc}$ in \eq{kappa}, unlike the direct tunneling,
can take positive, negative, or even complex values. 

The explicit expression for the Rabi frequency ${\cal R}_{bc}$ depends on the
particular form of the em coupling between the internal levels
which can be one or multi-photon transition.
For a two-photon transition via an off-resonant intermediate level
$|e'\>$ (which for a laser can be electronically excited and for a 
rf/microwave an electronic ground state) we may adiabatically eliminate $|e'\>$
\cite{JAV95}. We then obtain ${\cal R}_{bc}= {{\cal E}_b{\cal E}^*_c d_{e'b}d_{e'c}^*/ (2\hbar^2\Delta)}$ and a contribution to the em-induced level shifts $-{|{\cal E}_j|^2 |d_{e'j}|^2/( 2\hbar^2\Delta)}$
for $|j\>$. Here $d_{e'j}\equiv\mathbf{d}_{e'j}\cdot \pol_j$ and we have assumed that $|b\>$ 
is coupled to $|e'\>$ by the em field with the positive
frequency component ${\bf E}^+_b = \half
\pol_b{\cal E}_b e^{i {\bf k}_{b}\cdot \rv } e^{i\Omega_b t}$ and
detuning $\Delta$. 
Using the Gaussian approximation to $\phi_{bn}$, we obtain
$\kappa_{jk}\simeq \hbar {\cal R}_{jk} \epsilon_{jk} \exp{[-(3+\sqrt{3})\pi^2\sqrt{\bar{s}_{jk}}/48\sqrt{2}]}$,
for $(j,k)=(1,2),(2,3)$ and $\kappa_{13}\simeq \hbar {\cal R}_{13}
\epsilon_{13} \exp{(-\pi^2\sqrt{\bar{s}_{13}}/4\sqrt{6})}$, where
$\bar{s}_{jk}\equiv 4 s_j s_k/(\sqrt{s_j}+\sqrt{s_k})^2$ and 
$\epsilon_{jk}\equiv[\bar{s}_{jk}^2/(s_j s_k)]^{1/4}$. Here we have
assumed that the lattice potentials by the lasers $\alpha$ and $\beta$ in \eq{potential}
have the same amplitude in each level $|j\>$, i.e.\  $s_{j\alpha}=s_{j\beta}$,
and we have suppressed in the notation the index referring to the particular laser.

The onsite interaction term may also be obtained from the Wannier
functions $U_{jj}\simeq g_{2D}^{(jj)}\int dy\,dz\,|\phi_{j n}|^4$. 
The 2D nonlinearity 
$g_{2D}^{(jk)}\simeq 2\pi\hbar^2 a_{jk}/m l_x \sqrt{2\pi}$ is given 
in terms of the scattering length $a_{jk}$ and the 2D trap 
confinement $l_x=\sqrt{\hbar/m\omega_x}$ where 
the oscillator frequency perpendicular to the lattice is $\omega_x$. 
The additional density-dependent
contribution to the scattering length in 2D is negligible when
$\sqrt{2\pi}l_x/a_{jj}\gg \ln{(8\pi^{3/2}l_x n_{2D} a_{jj})}$ \cite{PET00},
where $n_{2D}$ denotes the 2D atom density.
For fermionic atoms $a_{jj}=0$ and the onsite interaction term vanishes for a single-species gas if the atoms 
only occupy the lowest mode of each lattice site, but is non-vanishing for a two-species gas trapped
in each lattice site. The nn interaction due to the $s$-wave scattering between the 
atoms in different sublevels is always non-vanishing for both bosonic and fermionic atoms.
It is generated in the spatial overlapping area of the adjacent sites and may be estimated 
by $U_{jk}^{\rm nn}\simeq 2g_{2D}^{(jk)}\int dy\,dz\,|\phi_{j n}|^2 |\phi_{k n'}|^2$.
We obtain
$U_{jj}\simeq  a_{jj} 3^{1/4} E_r \sqrt{s_j}/ l_x \sqrt{\pi}$, $U_{13}^{\rm nn}\simeq 2  a_{13} 3^{1/4} E_r\sqrt{\bar{s}_{13}}\, e^{-\pi^2\sqrt{\bar{s}_{13}}/2\sqrt{6}} / l_x\sqrt{\pi}$, and for other states
$U_{jk}^{\rm nn}\simeq 2  a_{jk} 3^{1/4}E_r \sqrt{\bar{s}_{jk}}\,
e^{-(3+\sqrt{3})\pi^2\sqrt{\bar{s}_{jk}}/24\sqrt{2}} / l_x\sqrt{\pi}$.

The Hamiltonian for the atomic system then reads
\begin{align}
    H &= \sum_k \left[\epsilon_k c^\dagger_k c_k + {U_{kk}} c^\dagger_k c^\dagger_k
c_k c_k \right]-\sum_{\<jk\>'}(J_{b}^{\bf p} c^\dagger_{j}c_k + {\rm H.c.})\nonumber\\
& -\sum_{\<jk\>}\left[(\kappa_{bc} c^\dagger_{j}c_k+{\rm H.c.}) +U_{bc}^{\rm nn} c^\dagger_j c_j c^\dagger_{k}
c_{k}\right],\label{ham}
\end{align}
where $\<jk\>$ denotes the summation over adjacent lattice sites and $\<jk\>'$ over the nearest
sites that are occupied by the same atomic species. The level shifts and the detunings due to
em-induced hoppings are included in $\epsilon_k$. Here $J_{b}^{\bf p}>0$ but all other coefficients
can take either positive or negative values, and $\kappa_{bc}$ can be complex.

We may compare the different terms in the Hamiltonian. For typical
experimental parameters for bosonic atoms we have $U_{jj}\gg U_{jk}^{\rm nn}$ ($j\neq k$).
For simplicity, assuming $s_1=s_2$, we obtain for lattice
heights $s=25$ and $40$, $U_{12}^{\rm nn}/U_{jj}\simeq 10^{-3}a_{12}/a_{jj}$ 
and $10^{-4}a_{12}/a_{jj}$. At the same lattice heights the em-driven hopping amplitudes 
in terms of the direct tunneling and the nn interactions are given by
$\kappa_{12}/J_1^{\kv_\alpha}\simeq 540 (\hbar {\cal R}_{12}/E_r)$
and $5100 (\hbar {\cal R}_{12}/E_r)$, and $\kappa_{12}/U_{12}^{\rm nn}
\simeq 4.2 (l_x/a_{12})(\hbar {\cal R}_{12}/E_r)$ and $8.3 (l_x/a_{12})(\hbar {\cal
R}_{12}/E_r)$. In shallow lattices, with weak em-coupling between the
sublevels $\hbar {\cal R}_{12}/E_r\ll 1$, the nn
hopping $\kappa_{bc}$ and the direct tunneling $J_b^{\bf p}$ between more distant
sites may be comparable. In deeper lattices and for stronger $\kappa_{bc}$ the direct tunneling
terms may be ignored. If the transverse confinement of the 2D lattice is weak ($l_x$ large) and
the Rabi field sufficiently strong $\hbar {\cal R}_{12} l_x/(a_{12}E_r)\gg 1$,
we may also neglect $U_{bc}^{\rm nn}$ with $\kappa_{bc}\gg U_{bc}^{\rm nn}$
and in \eq{ham} we only keep terms proportional to $\epsilon_k$, 
$U_{kk}$, and $\kappa_{bc}$. An especially interesting property of the proposed scheme
is that we may also find a wide range of parameter values for which $U_{bc}^{\rm nn} 
\agt \kappa_{bc}$. This limit can always be achieved with sufficiently weak em coupling.
If we also simultaneously require that $\kappa_{bc}\gg J_b^{\bf p}$, we may, e.g., at
$s=50$ select $\hbar {\cal R}_{12}/E_r\simeq 5\times 10^{-4}$ and $l_x\simeq 17 a_{12}$, resulting
in $ U_{bc}^{\rm nn} \simeq 10 \kappa_{bc}$. For instance, for $^{87}$Rb the $s$-wave
scattering length between $|F=1,M_F=-1\>$ and $|2,+1\>$ hyperfine states
is about 5.191nm \cite{HAR02}, corresponding to the transverse trap frequency of $\omega_x
\simeq 2\pi\times 15$~kHz, achievable, e.g., by an optical lattice.
The interspecies scattering length could be increased
by Feshbach resonances, further enhancing the effect of the nn interactions 
$U_{bc}^{\rm nn}$, so that one could reach $U_{jj}\gg U_{bc}^{\rm nn} \gg \kappa_{bc}\gg J_b^{\bf p}$
also in shallow lattices. Interspecies Feshbach resonances were
observed also in $^{87}$Rb between different hyperfine states \cite{feshbach} and
in optical lattices at low occupation numbers the harmful three-body losses are suppressed. 
Moreover, alkaline-earth-metal and rare-earth-metal atoms
with very narrow optical resonances \cite{yb} may allow the lattice lasers to be tuned
close to the atomic resonance making it easier to produce deep lattices.

The simplest situation is to select the phases of the em fields in \eq{kappa} so
that all the hopping amplitudes $\kappa_{bc}$ are real and positive. 
It is, however, also possible to engineer
a non-uniform phase profile for the hopping amplitudes which was in Ref.~\cite{RUO02} proposed
as a mechanism to construct topologically non-trivial ground states with fractional fermion numbers 
in 1D. In a 2D lattice the technique can
be used to create an effective magnetic field for neutral atoms \cite{JAK03}. Here we may similarly 
induce a phase for atoms hopping around a closed path in the lattice,
mimicing a magnetic flux experienced by charged particles. We write $\kappa_{bc}=|\kappa_{bc}|e^{i\nu_{bc}}$
where the phases $\nu_{bc}(\rv)$ may be constant or spatially varying \cite{RUO02,JAK03}. The hopping
around one unit triangle then generates the phase $\Delta\nu=\nu_{12}+\nu_{23}-\nu_{13}$ for the atoms,
corresponding to a magnetic flux $\Phi\propto\Delta\nu$ through the area enclosed by the triangle. 
For instance, for spatially constant $\nu_{bc}$ with $\Delta\nu\neq 0$, the entire
lattice area may be divided into side-sharing triangles where each adjacent triangle experiences
the flux with the opposite sign. In the case of atoms occupying more than one sublevel in each 
lattice site, we may also generate non-Abelian vector potentials similarly to Ref.~\cite{OST05};
see also \cite{WIL84,RUS05}.

For strong nn interactions with 1/3 filling, even without a vector potential, our model 
produces a frustrated ground state where
one atom in each triangle of sites is strongly influenced by the atoms in other corner-sharing triangles.
It is helpful to consider a honeycomb lattice, formed by connecting the centers of triangles of the 
Kagom\'e lattice, where the sites are coupled by ring-hopping processes \cite{POL06}.
A lattice system described by an analogous quantum dimer model on a pyrochlore/checkerboard 
lattice and on a 3D diamond lattice was recently shown to support fractional charges \cite{FULDE}. 
Similar fractional excitations are expected to exist in the Kagom\'e lattice system, where they 
can act as independent, deconfined particles over finite distances at temperatures above the ordering 
transition driven by quantum fluctuations -- in this case by the ring-exchange processes \cite{nic}.
Atomic states in the prepared lattice system could potentially be 
detected optically \cite{deb}.

Our formalism considers one atomic species per lattice site. It is straightforward to generalize it
to the situation where a two-species gas is trapped in each site with
em field inducing hopping for both species. Then the onsite interaction and the hopping
terms can be expressed as an effective Heisenberg spin-1/2 Hamiltonian $H_{\rm eff} \simeq \sum_{\<i,j\>}
[t_z S^z_i S^z_j\pm t_\perp (S^x_i S^x_j +S^y_i S^y_j)]$ where the $+$ ($-$) sign refers to fermionic
(bosonic) atoms and $S^k_i$ denote the spin matrices \cite{DUA03}. The
fermionic version has been extensively studied in 2D Kagom\'e lattices where the ground state of the SU(2) symmetric case ($t_z=t_\perp$) still has unsettled questions, e.g., in 
the existence of spontaneously broken symmetries and finite energy gaps \cite{spinkagome}.

We are grateful to N.\ Shannon for explanations of
the importance of Kagom\'e lattices and EPSRC for funding.

\end{document}